\documentclass{PoS}
\usepackage{amsmath}

\PoS{PoS(LAT2005)235}

\title{The Schr\"odinger functional with chirally 
rotated boundary conditions\footnote{FTUAM-05-13 and IFT-UAM-CSIC/05-39}}

\ShortTitle{The Schr\"odinger functional with chirally rotated boundary conditions}

\author{\speaker{Stefan Sint}\\
        Universidad Aut\'onoma de Madrid\\
	Instituto de F\'{\i}sica Te\'orica CSIC/UAM\\
	E-28049 Cantoblanco, Madrid, Spain\\
        E-mail: \email{stefan.sint@uam.es}}

\abstract{Using orbifold techniques I construct the 
Schr\"odinger functional (SF) for a doublet of Wilson quarks
with chirally rotated boundary conditions.
This allows to perform checks of universality: for instance,
the renormalized SF coupling constant, defined 
with either boundary conditions, must have a unique continuum limit.
Similarly, SF correlation functions in twisted mass QCD
and standard QCD can be defined such that they share a common
continuum limit. An additional benefit of the new set-up consists
in the observation that all the bulk O($a$) counterterms to the
action and composite operators become irrelevant in the chiral limit.
This implies that (ratios of) SF renormalization constants
can be automatically O(a) improved, up to the effect of unavoidable 
boundary counterterms.
As a first application we calculate the running coupling for $N_{\rm f}=2$ 
flavours in the SF-scheme to one-loop order of perturbation theory.
Universality of the continuum limit is confirmed and
the irrelevance of the Sheikholeslami-Wohlert term in the action
is demonstrated explicitly.}

\FullConference{XXIIIrd International Symposium on Lattice Field Theory\\
		 25-30 July 2005\\
		 Trinity College, Dublin, Ireland}

\begin{document}

\section{Introduction}
The Schr\"odinger functional (SF) has become a general tool
to address non-perturbative renormalization problems
in QCD\footnote{see \cite{Jansen:1995ck,Sint:2000vc} for an overview and
\cite{Guagnelli:2005zc} for a recent application and further references.}. 
Renormalization schemes based on the SF are gauge
invariant, quark mass independent (through renormalization in the 
chiral limit) and suitable for evaluation by both Monte Carlo and 
perturbative methods. Moreover, the finite space-time volume is used
to set the renormalization scale, so that recursive finite
size techniques become applicable.

Why should one be interested in changing the boundary conditions
for the quark and anti-quark fields?
The first motivation comes from twisted mass QCD: as 
discussed in~\cite{Frezzotti:2000nk}, correlation
functions calculated in renormalized twisted mass QCD
are, up to cutoff effects, related to renormalized correlation 
functions in standard QCD by a non-singlet chiral rotation.
For this statement to hold true with SF correlation functions,
the SF boundary conditions must be chirally rotated, too.
Formulating the SF with the same boundary conditions for both
twisted mass QCD and standard QCD (as was done in~\cite{Frezzotti:2001ea}) 
implies that the renormalised SF correlation functions 
are different even in the continuum limit.
Second, it can be shown that, in a finite volume 
with (some variant of) periodic boundary conditions, 
O($a$) improvement of massless Wilson quarks is automatic 
(cp.~\cite{Frezzotti:2003ni}). 
While the argument does not go through in the presence of
standard SF boundary conditions, it can be resurrected 
in the chirally rotated set-up.

\section{Chiral rotation and SF boundary conditions}

Consider isospin doublets $\chi'$ and $\overline{\chi}'$ 
of quark and anti-quark fields satisfying homogeneous 
SF boundary conditions~\cite{Sint:1993un} ($P_\pm = \frac12(1\pm\gamma_0$),
\begin{xalignat}{2}
     P_+\chi'(x)\vert_{x_0=0} &= 0,    &P_-\chi'(x)\vert_{x_0=T} &=0,
   \nonumber\\
   \overline{\chi}'(x)P_-\vert_{x_0=0} &= 0,   
    &\overline{\chi}'(x) P_+\vert_{x_0=T} &= 0. 
\end{xalignat}
When performing a chiral field rotation,
\begin{equation}
    \chi'=\exp(i\alpha \gamma_5\tau^3/2)\chi,\qquad
    \overline{\chi}'=\overline{\chi} \exp(i\alpha \gamma_5\tau^3/2),
\end{equation}
the rotated fields satisfy the chirally rotated boundary conditions
\begin{xalignat}{2}
     P_+(\alpha)\chi(x)\vert_{x_0=0} &=0,    
    &P_-(\alpha)\chi(x)\vert_{x_0=T} &=0,\nonumber\\
   \overline{\chi}(x)\gamma_0P_-(\alpha)\vert_{x_0=0} &= 0,   
    &\overline{\chi}(x)\gamma_0P_+(\alpha)\vert_{x_0=T} &= 0, 
\label{eq:rotbc}
\end{xalignat}
with the projectors
\begin{equation}
  P_\pm(\alpha)=\frac12\left[1\pm\gamma_0\exp(i\alpha\gamma_5\tau^3)\right].
\end{equation}
For $\alpha=0$ the standard projectors $P_\pm=P_\pm(0)$ are recovered.
The special case of $\alpha=\pi/2$ will be of 
particular interest in the following,
\begin{equation}
   P_\pm(\pi/2)\equiv Q_\pm=\frac12(1\pm i\gamma_0\gamma_5\tau^3),
\end{equation}
as the orbifold method cannot be applied for arbitrary
values of $\alpha$.

\section{Orbifold construction}
In the context of lattice QCD orbifold techniques have been 
applied by Taniguchi~\cite{Taniguchi:2004gf} in order to implement SF boundary
conditions for Ginsparg-Wilson quarks. 
Here I consider a single Wilson quark flavour with lattice action
\begin{equation}
  S_f[\psi,\bar\psi,U] = a^4\!\!\!\sum_{-T< x_0\leq T}
  \sum_{\bf x}\bar\psi(x)\left(D_W+m_0\right)\psi(x).
\end{equation}
The fermionic fields are taken to be $2T$-anti-periodic in the Euclidean time
direction,
\begin{equation}
   \psi(x_0+2T,{\bf x})=-\psi(x),\qquad
   \overline{\psi}(x_0+2T,{\bf x})=-\overline{\psi}(x),
\end{equation}
and an orbifold reflection is introduced about the point $x_0=0$,
\begin{equation}
   R: \psi(x) \rightarrow
        i\gamma_0\gamma_5\psi(-x_0,{\bf x}),
\qquad
      \overline{\psi}(x) \rightarrow
      \overline{\psi}(-x_0,{\bf x})i\gamma_0\gamma_5.
\end{equation}
Following~\cite{Taniguchi:2004gf} the gauge field can be treated as
an external field. It is first extended to the doubled time interval
$[-T,T]$, through
\begin{equation}
   U_k(-x_0,{\bf x})=U_k(x_0,{\bf x}),\qquad 
   U_0(-x_0-a,{\bf x})^\dagger=U_0(x),
\end{equation}
and then $2T$-periodically continued to all Euclidean times.

The fermionic fields may now be decomposed in even and odd
components with respect to the reflection symmetry $R$,
viz.
\begin{equation}
  R\psi_\pm= \pm\psi_\pm,\qquad R\overline{\psi}_\pm = \pm \overline{\psi}_\pm.
\end{equation}
Even and odd fields then satisfy Dirichlet conditions at $x_0=0$,
\begin{equation}
 (1\mp i\gamma_0\gamma_5)\psi_\pm(0,{\bf x})=0,
 \qquad \overline{\psi}_\pm(0,{\bf x})(1\mp i\gamma_0\gamma_5) = 0.
\end{equation}
Due to the $2T$-anti-periodicity the complementary components
then satisfy Dirichlet conditions at $x_0=T$:
\begin{equation}
  (1\pm i\gamma_0\gamma_5)\psi_\pm(T,{\bf x})=0,
  \qquad \overline{\psi}_\pm(T,{\bf x})(1\pm i\gamma_0\gamma_5) = 0.
\end{equation}
For the orbifold projection to work, a consistency condition
must be satisfied,
\begin{equation}
  S_f[\psi,\overline{\psi},U] =
  S_f[\psi_++\psi_-,\overline{\psi}_++\overline{\psi}_-,U]
  = S_f[\psi_+,\overline{\psi}_+,U] + S_f[\psi_-,\overline{\psi}_-,U].
\end{equation}
This condition is rather strong and is the reason why the
construction is restricted to the choice $\alpha=\pi/2$.
As a consequence, the functional
integral over the fermion fields factorises.
Interpreting $R$-even and $R$-odd fields as flavour components
of a doublet field,
\begin{equation}
  \chi=\sqrt{2}
  {\begin{pmatrix}\psi_-\\ \psi_+ \end{pmatrix}},\qquad
  \overline{\chi}= \sqrt{2}
  {\begin{pmatrix}\overline{\psi}_-\,, & \overline{\psi}_+\end{pmatrix}},
\label{eq:doublet}
\end{equation}
the functional integral over the quark fields takes the form
\begin{equation}
  \label{eq:factorize}
 \int \prod_{-T < x_0 \le T} {\rm d}\psi(x){\rm d}\overline{\psi}(x)
  {\rm e}^{-S_f[\psi,\overline{\psi},U]} \propto
  \int\prod_{0\le x_0 \le T} 
  {\rm d}\chi(x){\rm d}\overline{\chi}(x) 
  {\rm e}^{-\frac12 S_f[\chi,\overline{\chi},U]}.
\end{equation}
The fields in the orbifolded theory satisfy the boundary conditions
\begin{xalignat}{2}
     Q_+\chi(x)\vert_{x_0=0} &= 0,    &Q_-\chi(x)\vert_{x_0=T} &=0,\nonumber\\
   \bar\chi(x)Q_+\vert_{x_0=0} &= 0,   &\bar\chi(x) Q_-\vert_{x_0=T} &= 0, 
\end{xalignat}
which are a special case of Eqs.~(\ref{eq:rotbc}).
Furthermore, in Eq.~(\ref{eq:factorize}) we have used the fact 
that the independent field variables $\chi$ and $\overline{\chi}$
may be taken to be the fields at Euclidean times  $0<x_0<T$ 
and the non-Dirichlet components at $x_0=0,T$.
Finally, a complete reduction to the interval $[0,T]$ is obtained
by  re-writing the lattice action
\begin{equation}
   S_f[\chi,\overline{\chi},U] = 
   2a^4\!\!\! \sum_{0\le x_0 \le T} 
 \sum_{\bf x}\overline{\chi}(x){\cal D}\chi(x).
\end{equation}
Here ${\cal D}$ is essentially the Wilson-Dirac operator 
including the standard bare mass term, except for some modifications
near the time boundaries which are induced by the orbifold
reflection. The factor of 2 originates from the fact
that the contributions from negative and positive times are equal,
and has been anticipated in Eq.~(\ref{eq:doublet}).
The explicit form of ${\cal D}$ and the 
identification of the dynamical field space 
are the main results of the orbifold construction which
would have been difficult to obtain in other ways.

\section{Boundary counterterms}

Using the lattice symmetries of the chirally rotated SF
the possible boundary counterterms of dimension 3 
can be identified:
\begin{equation}
    K_1=\overline{\chi}\gamma_5\tau^3\chi,\qquad 
  K_\pm=\overline{\chi}Q_\pm\chi.
\end{equation}
It is easy to see that $K_1$ corresponds to the
logarithmically divergent boundary counterterm
in the standard SF~\cite{Sint:1995rb}.
It leads to a multiplicative renormalization of
the quark and anti-quark boundary fields.
In order to understand the r\^ole of $K_\pm$
it is instructive to rotate back to the standard
SF. In terms of the primed fields one then obtains
\begin{equation}
   K_\pm \rightarrow -\overline{\chi}'i\gamma_5\tau^3P_\pm\chi'.
\end{equation}
We note that $K_+$ ($K_-$) contains only Dirichlet components
at $x_0=0$ ($x_0=T$). This means that it will never contribute
to the SF correlation functions used in practice. 
We are thus left with a single counterterm $K_-$ at $x_0=0$ 
($K_+$ at $x_0=T$), which is composed of non-Dirichlet
components only. As it violates parity and flavour
symmetries it is a {\em finite} counterterm which can 
be fixed by requiring parity restoration.
The extension of this analysis to
dimension 4 boundary counterterms
will be discussed elsewhere~\cite{Sintprep}.

\section{O($a$) improvement and SF boundary conditions}

Consider first massless lattice QCD on a hyper-torus with some
variant of periodic boundary conditions for all fields.
The cutoff dependence of renormalized correlation functions 
is then described by Symanzik's 
effective continuum theory~(see \cite{Luscher:1996sc} for 
notation and references),
\begin{eqnarray}
   S_{\rm eff} &=& S_0+ a S_1 + O(a^2),\nonumber\\
  \langle O\rangle &=&
  \langle O\rangle^{\rm cont}
  +a \langle S_1 O\rangle^{\rm cont}
  +a \langle \delta O\rangle^{\rm cont} + O(a^2).
\end{eqnarray}
Here, $S_1$ and $\delta O$ are O($a$) counterterms for the action and for 
$O$. Chiral symmetry of the continuum action $S_0$ implies that
all insertions of O($a$) counterterms vanish. To see this more 
explicitly consider a $\gamma_5$ field transformation:
\begin{equation}
   \chi\rightarrow\gamma_5\chi,\qquad 
   \overline{\chi}\rightarrow -\overline{\chi}\gamma_5.
\end{equation}
The massless continuum action $S_0$ is invariant
while $S_1$ changes its sign. Assuming that $O$ is even under
a $\gamma_5$ transformation, one can also show that $\delta O$ 
has to be odd. It then follows that
\begin{eqnarray}
  \langle O S_1\rangle^{\rm cont} = - \langle O S_1\rangle^{\rm cont},
  \qquad 
   \langle \delta O \rangle^{\rm cont} = 
  -\langle \delta O \rangle^{\rm cont},
\end{eqnarray}
implying that indeed both counterterm insertions vanish. 
One may wonder whether the O($a$) ambiguity of the chiral
limit might affect this conclusion. This is
not the case, as the ambiguity is proportional to  
\begin{equation}
  \int{\rm d}^4 x\langle O \overline{\chi}(x)\chi(x)\rangle^{\rm cont} = 0,
\end{equation}
which vanishes again due to chiral symmetry.
It follows that O($a$) improvement of massless Wilson quarks 
in finite volume is automatically satisfied~(cp.~\cite{Frezzotti:2003ni}).

The situation changes in the presence of SF boundary conditions.
The $\gamma_5$ transformation is then no longer a symmetry 
of the effective continuum theory, as the transformed fields
satisfy boundary conditions with the complementary projectors.
Indexing correlation functions with the projectors of the
boundary condition for $\chi$ at $x_0$,
one finds for $\gamma_5$-even observables $O$,
\begin{equation}
  \langle O\rangle^{\rm cont}_{(P_\pm)} =
   \langle O\rangle_{(P_\mp)}^{\rm cont},\qquad
   \langle O S_1\rangle^{\rm cont}_{(P_\pm)} =
   -\langle O S_1\rangle_{(P_\mp)}^{\rm cont}\neq 0,
\end{equation}
i.e.~the insertion of $S_1$ does not vanish.
This should not come as a surprise, as otherwise the
determination of O($a$) improvement coefficients
such as $c_{\rm sw}$ or $c_{\rm A}$ in ref.~\cite{Luscher:1996ug}
would have been impossible.

The question then arises whether automatic O($a$) improvement
can be achieved with the chirally rotated boundary 
conditions\footnote{For a different attempt see~\cite{Frezzotti:2005zm}.}.
This is indeed the case: if one augments the 
$\gamma_5$ transformation by a flavour exchange, i.e.
\begin{equation}
   \chi\rightarrow\gamma_5\tau^1\chi,\qquad 
   \overline{\chi}\rightarrow -\overline{\chi}\gamma_5\tau^1,
\end{equation}
the chirally rotated SF boundary conditions remain unchanged,
due to $\gamma_5\tau^1 Q_\pm = Q_\pm \gamma_5\tau^1$.
Therefore, for $\gamma_5\tau^1$-even observables
the previous argument holds. Note, however, that this discussion
does not account for the O($a$) counterterms at the
time boundaries. Some of them share symmetries with the
continuum action $S_0$ and their insertion therefore 
never cancels. However, O($a$) improvement may 
then still be achieved by tuning just a few O($a$) boundary counterterms.

The SF coupling~\cite{Luscher:1992an} 
is an example of a $\gamma_5$-even observable.
In perturbation theory one finds
\begin{equation}
  \bar{g}^2(L)=g^2_{\overline{\rm MS}}(\mu)+ k_1(\mu L)
 g^4_{\overline{\rm MS}}(\mu) + O(g^6).
\end{equation}
Setting $\mu=L^{-1}$ the fermionic contribution
to  $k_1=k_{1,0}+N_{\rm f} k_{1,1}$
has been computed in~\cite{Sint:1995ch},
\begin{equation}
   k_{1,1}= -0.039863(2)/(4\pi),
\end{equation}
and is obtained as the leading term in the 
asymptotic behaviour of a series of lattice approximants,
\begin{equation}
 f(L/a)\sim r_0 + (a/L)\left[r_1+s_1\ln(a/L)\right]+O(a^2).
\end{equation}
In the standard SF one finds 
$r_0=k_{1,1}$ and $s_1\propto c_{\rm sw}^{(0)}-1$. 
Using instead the chirally rotated SF, $k_{1,1}$ is indeed reproduced,
while $s_1$ now vanishes {\em independently} of $c_{\rm sw}^{(0)}$,
as expected.

%

\subsection*{Acknowledgments}
I would like to thank Margarita Garc\'{\i}a P\'erez 
and Rainer Sommer for useful discussions. 
Partial support by the Spanish government
through a Ram\'on y Cajal fellowship is gratefully acknowledged.


\begin{thebibliography}{99}

\bibitem{Jansen:1995ck}
  K.~Jansen {\it et al.},
  ``Non-perturbative renormalization of lattice QCD at all scales,''
  Phys.\ Lett.\ B {\bf 372} (1996) 275
  [arXiv:hep-lat/9512009].

\bibitem{Sint:2000vc}
  S.~Sint,
  ``Non-perturbative renormalization in lattice field theory,''
  Nucl.\ Phys.\ Proc.\ Suppl.\  {\bf 94} (2001) 79
  [arXiv:hep-lat/0011081].

\bibitem{Guagnelli:2005zc}
  M.~Guagnelli, J.~Heitger, C.~Pena, S.~Sint and A.~Vladikas  [ALPHA
                  Collaboration],
  ``Non-perturbative renormalization of left-left four-fermion operators in
  quenched lattice QCD,''
  arXiv:hep-lat/0505002.

\bibitem{Frezzotti:2000nk}
  R.~Frezzotti, P.~A.~Grassi, S.~Sint and P.~Weisz  [Alpha collaboration],
  ``Lattice QCD with a chirally twisted mass term,''
  JHEP {\bf 0108} (2001) 058
  [arXiv:hep-lat/0101001].

\bibitem{Frezzotti:2001ea}
  R.~Frezzotti, S.~Sint and P.~Weisz  [ALPHA collaboration],
  ``O(a) improved twisted mass lattice QCD,''
  JHEP {\bf 0107} (2001) 048
  [arXiv:hep-lat/0104014].

\bibitem{Frezzotti:2003ni}
  R.~Frezzotti and G.~C.~Rossi,
  ``Chirally improving Wilson fermions. I: O(a) improvement,''
  JHEP {\bf 0408} (2004) 007
  [arXiv:hep-lat/0306014].

\bibitem{Sint:1993un}
  S.~Sint,
  ``On the Schr\"odinger functional in QCD,''
  Nucl.\ Phys.\ B {\bf 421} (1994) 135
  [arXiv:hep-lat/9312079].

\bibitem{Taniguchi:2004gf}
  Y.~Taniguchi,
  ``Schr\"odinger functional formalism with Ginsparg-Wilson fermion,''
  arXiv:hep-lat/0412024.

\bibitem{Sint:1995rb}
  S.~Sint,
  ``One loop renormalization of the QCD Schr\"odinger functional,''
  Nucl.\ Phys.\ B {\bf 451} (1995) 416
  [arXiv:hep-lat/9504005].

\bibitem{Sintprep}
  S.~Sint, in preparation

\bibitem{Luscher:1996sc}
  M.~L\"uscher, S.~Sint, R.~Sommer and P.~Weisz,
  ``Chiral symmetry and O(a) improvement in lattice QCD,''
  Nucl.\ Phys.\ B {\bf 478} (1996) 365
  [arXiv:hep-lat/9605038].

\bibitem{Luscher:1996ug}
  M.~L\"uscher, S.~Sint, R.~Sommer, P.~Weisz and U.~Wolff,
  ``Non-perturbative O(a) improvement of lattice QCD,''
  Nucl.\ Phys.\ B {\bf 491} (1997) 323
  [arXiv:hep-lat/9609035].

\bibitem{Frezzotti:2005zm}
  R.~Frezzotti and G.~Rossi,
  ``Chirally improving Wilson fermions. III: The Schr\"odinger functional,''
  arXiv:hep-lat/0507030.

\bibitem{Luscher:1992an}
  M.~L\"uscher, R.~Narayanan, P.~Weisz and U.~Wolff,
  ``The Schr\"odinger functional: A renormalizable probe for non-Abelian gauge
  theories,''
  Nucl.\ Phys.\ B {\bf 384}, 168 (1992)
  [arXiv:hep-lat/9207009].

\bibitem{Sint:1995ch}
  S.~Sint and R.~Sommer,
  ``The Running coupling from the QCD Schr\"odinger functional: A One loop
  analysis,''
  Nucl.\ Phys.\ B {\bf 465} (1996) 71
  [arXiv:hep-lat/9508012].

\end{thebibliography}
\end{document}